\documentstyle[aps,prb,multicol]{revtex}
\def\eps{\epsilon}

\begin{document}
%\preprint{2001-}
\title{Condensate  of  low dimensional 
charged Bose disks in a uniform magnetic field}
\author{Sang-Hoon Kim}
\address{Division of Liberal Arts, Mokpo National Maritime University,
 Mokpo 530-729, Korea}
\author{Peter Pfeifer}
\address{Department of Physics and Astronomy,
 University of Missouri-Columbia, 
Columbia, MO 65211}
\date{\today}
\maketitle
\draft
\begin{abstract}
The Bose condensation of a stack of low dimensional disks
which is composed of a noninteracting charged Bose gas
in a uniform magnetic field is studied.
A statistical approach with density of states at noninteger dimensions
are applied for the system.
The condensate fraction of the disk system
 in a uniform magnetic field is calculated.
The stack of low dimensional charged Bose disks is found not to share 
the  condensate  behavior of the traditional BCS superfluids.
\end{abstract} 
%\newpage
\begin{multicols}{2}

%%%%%%%%%%%%%%%%%%%%%%%%%%%%%%%%%%%%%%%%%%%%%%

It is known that two fermions can be coupled to form a
 pair which   behaves like a spinless boson.
Many bosonic pairs form a kind of charged Bose gas, and it has also been
 known that the condensation of the Bose system  could be 
a reliable candidate for superfluidity. \cite{alex4,alex1}
It has been reported that a dominant contribution to the superfluid 
density of liquid helium-4 in films and porous media originates from 
the geometric  structure, such as the noninteger dimensionality of the 
samples.\cite{kim} 
The dimensionality represents the effects in the measure of
disorderness in terms of the connectivity of the system.\cite{Pf}

We are interested in the superfluid density of a stack of noninteracting
 low dimensional charged Bose disks (CBD) in a uniform magnetic field.
We assume that the distance  between any two fermions of a pair
is large enough to neglect the Coulomb interaction.
Condensate density plays a key role in the superfluid density.
The low dimensional disk is defined as an object
 which has a dimension of between 1 and 2.
It could be a very thin, spotted system 
with negligible thickness to maintain the dimensionality less than 2.

However, a system which has a dimension of equal or less than 2
 can not produce  any nonzero superfluid density. 
It also has been known that an ideal charged Bose gas in two 
dimensions (2D) can not be condensed even under a magnetic field 
because of the one dimensional (1D) character of particle motion
within the lowest Landau level.\cite{scha,alex3} 
On the other hand, if we pile up the low dimensional disks in  parallel,
this gives an extra dimension to the perpendicular axis.\cite{hilf}
The whole dimension of the stack then
 becomes between 2 and 3, which  is large  enough to
 create a nonzero superfluid density. 
We apply  a simple statistical approach for the $D$ noninteger dimensions.

The theory we use begins from the noninteracting Bose gas 
in disk dimensions. It is uniform in disk directions.
 This is then extended to a charged system such as the
bipolaronic method for the condensate  density. \cite{alex4,alex1}
The  partition function of the system is given as
\begin{equation}
\ln {\mathcal{Q}}(z,T,D) = -\int_{0}^{\infty} d\eps \, \rho_D(\eps)
\ln (1-z e^{- \eps /T}) - \ln (1-z),
\label{11}
\end{equation}
where  $z$ is the fugacity defined by $z=e^{-|\mu|/T}$, 
and $\eps_{\bf p} = p^2/2m$ is taken for the neutral system.
The term $\rho_D(\eps)$ is the  density of states at  $D$-dimension,
the spectral dimension,\cite{Pf} 
and plays a key role in our analysis.
The spectral dimension of the stack of disks depends very much on 
how the disks are interconnected.

For a neutral and uniform system it is  given as \cite{kim,Pf}
\begin{equation}
\rho_D(\eps) = a_D \eps^{D/2 - 1},
\label{14}
\end{equation}
where $a_D$ is a $D$-dimensional coefficient given by  
$a_D =  \Gamma\left( \frac{D}{2} \right)^{-1}
\left( \frac{m^\ast}{2 \pi} \right)^{D/2}.$
Here, $\Gamma$ is the Gamma function and $m^\ast$ is
 the effective mass of a pair.
We set  $\hbar=c=k_B =1$ for convenience, and unit volume is assumed.

The average number of particles is obtained from Eq. (\ref{11})
\begin{eqnarray}
n &=& z\frac{\partial \ln {\mathcal{Q}}}{\partial z}
\nonumber \\
&=&\int_0^\infty d\eps \frac{\rho_D(\eps)}{z^{-1} (e^{\eps /T} - 1)}
+ n_0,
\label{15}\end{eqnarray}
 where $n_0 = n_{{\bf p}=0}$. 
Next, substituting the $D$-dimensional
density of states from Eq. (\ref{14}) into Eq. (\ref{15}),
the condensate fraction is obtained as\cite{kim,Pf,huan}
\begin{eqnarray}
\frac{n_0}{n} &=&
1- v \int_0^\infty d\eps \frac{\rho_D(\eps)}{z^{-1} (e^{\eps /T} - 1)}
\nonumber \\
&=& 1- \left\{ \int_0^\infty d\eps \frac{\eps^{D/2-1}}{e^{\eps /T} - 1}
\right\}_{T_c}^{-1}
\int_0^\infty d\eps \frac{\eps^{D/2-1}}{e^{\eps /T} - 1}
\nonumber \\
&=& 1-  \left\{ \frac{T}{T_c(D)} \right\}^{D/2},
\label{16}
\end{eqnarray}
where
\begin{equation}
T_{c}(D) = \frac{2 \pi }{m^\ast v^{2/3}}
\frac {1}{\zeta(\frac{D}{2})^{2/D}}.
\label{ctem}
\end{equation}
Here, $v$ is the volume density and $\zeta$ is the Riemann-Zeta function.
The $z=1$ limit is taken for the condensation.
Note that $\int_0^\infty dx \, x^{D/2 -1}/(e^x - 1) 
= \Gamma(D/2) \zeta(D/2).$

The critical temperature in Eq. (\ref{ctem}), $T_c$,  
 corresponds to the BEC transition temperature.
It is rewritten as a function of  $T_c^b$  for the bulk $(D=3)$
 as
\begin{equation}
T_c(D) = \frac{1.897}{\zeta\left(\frac{D}{2}\right)^{2/D}} T_c^b,
\label{40}
\end{equation}
where $\zeta(\frac{3}{2})^{2/3} = 1.897$.
 It can be readily shown that 
 Eq. (\ref{40}) satisfies both the ideal thin limit ($D=2$)
 and the bulk limit ($D=3$).
Note that the transition would not occur for the 2D limit since
 $ T_c  \sim  \left| \frac{D}{2}-1 \right|$
 as  $D$ approaches to 2.\cite{Grad}

The CBD  model in a uniform magnetic field is now extended
using this new $D$-dimensional density of states, $\rho_D(\eps,H)$.
Our $D$-dimensional system, $2<D<3$, is composed of $(D-1)$-dimensional
planes and an additional dimension which is parallel to 
the magnetic field.
A uniform magnetic field is applied  perpendicular to the direction
of the disks. 
The new density of states in $D$-dimensions is derived 
from the Landau quantization law.\cite{alex1,land}
	The $(D-1)$-dimensional degeneracy is
\begin{equation}
\rho_{D-1}(\eps) \omega_H = \frac{1}{\Gamma\left(\frac{D-1}{2}\right)}
\left( \frac{m^\ast}{2 \pi}\right)^{\frac{D-1}{2}}
\eps^{ \frac{D-1}{2} -1}\omega_H,
\label{43}
\end{equation}
where $\omega_H=2eH/m^\ast$ is the cyclotron frequency.
Therefore, $\rho_D(\eps,H)$ is obtained as
\begin{eqnarray}
\rho_D(\eps,H) &=& \rho_{D-1}(\eps) \omega_H
\sum_{n, p_z} \delta(\eps - \eps_{n,p_z})
\nonumber \\
&=& \frac{1}{\sqrt{\pi} \, \Gamma\left(\frac{D-1}{2}\right) }
\left( \frac{m^\ast}{2 \pi}\right)^{\frac{D}{2}}
\eps^{\frac{D-1}{2}}\omega_H
\nonumber \\ &\times&
\sum_{n=0}^\infty \frac{1}{ \sqrt{\eps - (n+\frac{1}{2})\omega_H}}.
\label{45}
\end{eqnarray}

The condensate fraction for the charged system is then obtained as
\begin{eqnarray}
\frac{n_0}{n}
&=& 1- \left\{ \int_0^\infty d\eps \frac{\rho_D(\eps,H)}{e^{ \eps /T} - 1}
 \right\}_{T_c}^{-1}
 \int_0^\infty d\eps \frac{\rho_D(\eps,H)}{e^{ \eps /T} - 1}
 \nonumber \\
&=& 1-  \left\{  \frac{T}{T_c(D)} \right\}^{D/2}
\frac{A(T,\omega_H)}{A(T_c,\omega_H)}.
\label{46}
\end{eqnarray}
$A(T,\omega_H)$  is defined by 
\begin{equation}
A(T,\omega_H)= \int_{x_0}^\infty dx \frac{x^{\frac{D-1}{2}}}{e^x - 1}\sum_{n=0}^\infty
\frac{1}{\sqrt{x-(n+\frac{1}{2})\frac{\omega_H}{T}}},
\label{47}
\end{equation}
where $x_0= (n+\frac{1}{2})\frac{\omega_H}{T}$.
Here, we are considering the range:  $\omega_H/T \ll 1$.
$A(T,\omega_H)$ itself can diverge, but $n_0/n$ in 
Eq. (\ref{46}) does not.

The condensate density of the charged Boson model contains
 one additional factor
 of $A$ which gives the effect of the field over that of 
the neutral model in Eq. (\ref{16}).
It is plotted in FIG. 1, at various disks,
as a function of normalized temperature $T/T_c^b$.
 The temperature $T_c$ is converted  with the help of Eq. (\ref{40}).

We find that the field strongly effects the condensate fraction of the 
noninteracting CBD.
Also, we see the higher the disk dimension, the larger the condensate 
fraction.
The most noticeable effect is that the condensate fraction increases as
the magnetic field is increased. 
This  means that the CBD of the low dimensional disks
does not share the condensate behavior with the conventional 
BCS superfluids of antiferromagnetism. 
This is not surpring because the system we have discussed 
is a noninteracting charged system, 
whereas real superconducting films are interacting charged systems.
Furthermore, the dimension of superconducting disk
 is  not low dimensional but quasi-two dimensional which is 
greater than 2.
However, if we consider a ferromagnetic material or superconductor
 which is composed of many disks, we suggest the
 possibility that it could be modeled by a composition
 of low dimensional disks of noninteracting bosonic pairs.

%%%%%%%%%%%%%%%%%%%%%%%%%%%%%%%%%%%%%%%%%%%%%%
We send special thank to Prof. A. S. Alexandrov,  C. K. Kim, and 
G. McIntosh for fruitful advice.

%%%%%%%%%%%%%%%%%%%%%%%%%%%%%%%%%%%%%%%%%%%%
%%%%%   REFERENCES   %%%%%%%%%
%%%%%%%%%%%%%%%%%%%%%%%%%%%%%%%%%%%%%%%%%%%%%%

%%%%%%%%%%%%%%%%%%%%%%%%%%%%%%%%%%%%%%%%%%%%%%
%%%%%%%%%%%%   Beginning of Figure caption    %%%%%%%%%%
%%%%%%%%%%%%%%%%%%%%%%%%%%%%%%%%%%%%%%%%%%%%%%
\begin{figure}
\caption{The condensate fraction  of the charged Bose disks in a uniform
magnetic field.
The solid line is for field free bosons, the dashed line is for
$\omega_H/T_c^b = 10^{-3}$, and the dotted line is for 
$\omega_H/T_c^b = 10^{-2}$.  
(a) When D=3.0 (the dimension of a disk is 2.0). 
(b) When D=2.6 (the dimension of a disk is 1.6).}
\end{figure}
%%%%%%%%%%%%%%%%%%%%%%%%%%%%%%%%%%%%%%%%%%%%%%
\end{multicols}
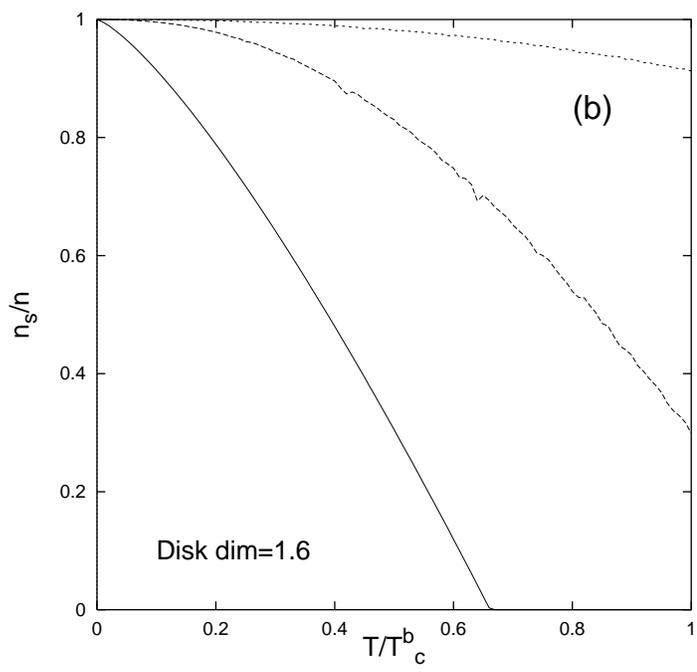
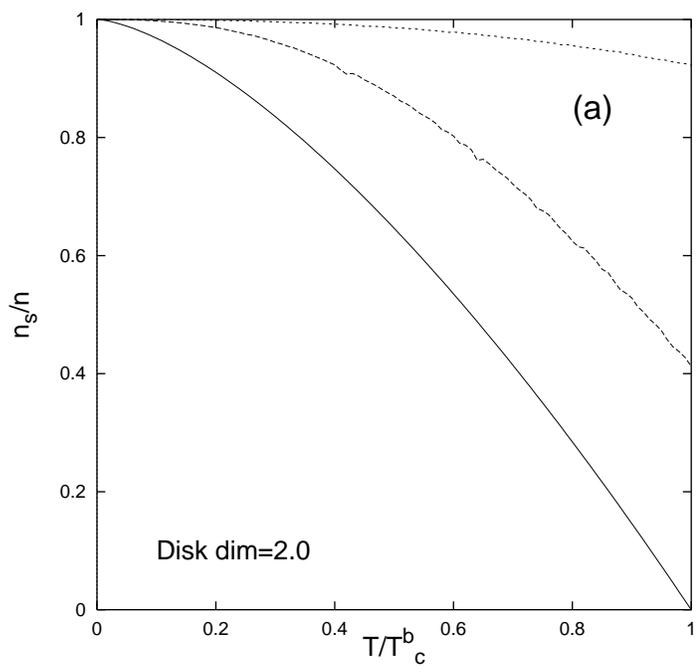
\end{document}